\documentclass[a4paper]{jpconf}
\usepackage{graphicx}
\begin{document}
\title{Current status of the T2K experiment
\footnote{Talk at \char'134 2nd International Conference on Particle Physics (ICPP-Istanbul II)",
Istanbul, Turkey, June 20-25, 2011.}
\footnote{The presentation file used in the talk can be found at
{\tt http://www-nu.kek.jp/}\~{\tt oyama/ICPP2011.t2k.oyama.final.ppt}}}

\thispagestyle{plain}
\pagestyle{plain}

\newcommand{\superk}           {Super-Kamiokande\xspace}       
\newcommand{\nue}                {$\nu_{e}$\xspace}
\newcommand{\nuebar}           {$\bar{\nu}_{e}$\xspace}
\newcommand{\numubar}           {$\bar{\nu}_{\mu}$\xspace}
\newcommand{\numu}             {$\nu_{\mu}$\xspace}
\newcommand{\nutau}             {$\nu_{\tau}$\xspace}
\newcommand{\nux}                {$\nu_{x}$\xspace}
\newcommand{\numunue}       {$\nu_{\mu} \rightarrow \nu_{e}$\xspace}
\newcommand{\numunux}       {$\nu_{\mu} \rightarrow \nu_{x}$\xspace}
\newcommand{\numunutau}    {$\nu_{\mu} \rightarrow \nu_{\tau}$\xspace}
\newcommand{\tonethree}       {$\theta_{13}$\xspace}
\newcommand{\tonetwo}         {$\theta_{12}$\xspace}
\newcommand{\ttwothree}       {$\theta_{23}$\xspace}
\newcommand{\ssttmue}          {$\sin^2 2 \theta_{{\mu}e}$\xspace}
\newcommand{\sstonethree}    {$\sin^2 2 \theta_{13}$\xspace}
\newcommand{\ssttwothree}    {$\sin^2 2 \theta_{23}$\xspace}
\newcommand{\msqonetwo}   {$\Delta m^2_{12}$\xspace}
\newcommand{\msqonethree} {$\Delta m^2_{13}$\xspace}
\newcommand{\msqtwothree} {$\Delta m^2_{23}$\xspace}
\newcommand{\absmsqtwothree} {$|\Delta m^2_{23}|$\xspace}
\newcommand{\msqmue}        {$\Delta m^2_{{\mu}e}$\xspace}
\newcommand{\msqmumu}     {$\Delta m^2_{\mu\mu}$\xspace}
\newcommand{\enu}               {$E_{\nu}$\xspace}
\newcommand{\pmu}              {$p_{\mu}$\xspace}
\newcommand{\amome}         {$E_{e}$\xspace}
\newcommand{\evis}              {$E_{vis}$\xspace}
\newcommand{\pizero}           {$\pi^{0}$\xspace}
\newcommand{\pizerogg}       {$\pi^{0}\to\gamma\gamma$\xspace}
\newcommand{\degree}      {$^\circ$\xspace}

\def\nue{\nu_{e}}
\def\num{\nu_{\mu}}
\def\nut{\nu_{\tau}}
\def\nmnt{$\nu_{\mu}\leftrightarrow\nu_{\tau}$~}
\def\nenm{$\nu_{e}\leftrightarrow\nu_{\mu}$~}
\def\nmne{$\nu_{\mu}\leftrightarrow\nu_{e}$~}
\def\lsim{\lower.7ex\hbox{${\buildrel < \over \sim}$}}
\def\gsim{\lower.7ex\hbox{${\buildrel > \over \sim}$}}
\author{Yuichi Oyama, on behalf of the T2K collaboration}

\address{KEK, Oho 1-1, Tsukuba Ibaraki 305-0801, Japan}

\ead{yuichi.oyama@kek.jp}

\begin{abstract}The T2K long-baseline neutrino-oscillation
experiment was started in January 2010 for the purpose of physics data-taking.
Until the massive earthquake on March 11, 2011 in Japan, $1.43\times 10^{20}$ pot data were accumulated.
In this data, 6 possible $\nu_{e}$ appearance candidates are found for
expected background of 1.5$\pm$0.3(syst.) if $\sin^{2}2\theta_{13}=0$ is assumed.
The probability that 6 or more events can be observed for
1.5$\pm$0.3 expected events is 7$\times$10$^{-3}$,
equivalent to 2.5$\sigma$ significance.
This is the first indication of a $\nu_{e}$ appearance or, in other words, 
the first indication of a non-zero $\sin^{2}2\theta_{13}$.
\end{abstract}

\section{Introduction}
In the presently known three-flavor framework of neutrinos, the flavor eigenstate
is a mixture of mass eigenstates:

$$
\left( \begin{array}{c}\nue \\ \num \\ \nut \end{array} \right) = {\rm{\bf U_{MNS}}}
\left( \begin{array}{c}\nu_{1} \\ \nu_{2} \\ \nu_{3} \end{array} \right).
$$

\noindent
The 3$\times$3 unitary mixing matrix, ${\rm {\bf U_{MNS}}}$, is known as 
the Maki-Nakagawa-Sataka (MNS) mass matrix\cite{MNS}. It has 6 independent parameters:
2 square mass differences ($\Delta m^{2}_{12}$ and $\Delta m^{2}_{23}$), 3 mixing angles
($\theta_{12}$, $\theta_{23}$, and $\theta_{13}$), and 1 CP-violating phase ($\delta_{\rm CP}$).
Among these parameters, $\Delta m^{2}_{12}$ and $\theta_{12}$ were determined
through solar\cite{SKsolar,SNO} and reactor\cite{Kamland} neutrino oscillation experiments:

$$ \Delta m^{2}_{12} \sim 8\times 10^{-5} {\rm eV^{2}~~~~~~~~~and~~~~~~~~~~} \theta_{12} \sim 34^{\circ}.$$

\noindent
On the other hand,  $\Delta m^{2}_{23}$ and $\theta_{23}$ were measured through atmospheric\cite{SKatm}
and long-baseline\cite{K2K,MINOSmu} neutrino-oscillation experiments: 

$$ \Delta m^{2}_{23} \sim 2.4\times 10^{-3} {\rm eV^{2}~~~~~~~~~and~~~~~~~~~~} \theta_{23} \sim 45^{\circ}.$$

\noindent
The remaining parameters, $\theta_{13}$ and $\delta_{\rm CP}$, are unknown.

The oscillation parameter $\theta_{13}$ can be determined by studying the following transition probability:

$$P(\nu_{\mu}\leftrightarrow\nu_{e}) \sim \sin^{2}2\theta_{13}\sin^{2}\theta_{23}\sin^{2}\Big({{\Delta m^{2}_{23}L}\over{4 E_{\nu}}}\Big)
\sim {{1}\over{2}}\sin^{2}2\theta_{13}\sin^{2}\Big({{\Delta m^{2}_{23}L}\over{4 E_{\nu}}}\Big).$$

\noindent
Two different types of experiments can be performed to determine non-zero $\theta_{13}$.
One is the $\nu_{e}$ disappearance experiment using $\sim$3MeV ${\bar \nu_{e}}$ from reactors
with baseline distance of a few kilometers.
At present, the upper limit on $\sin^{2}2\theta_{13}$ is known to be 0.15,
as reported from the CHOOZ\cite{CHOOZ} reactor experiment.
The other type of experiment is the $\nu_{e}$ appearance experiment
using an almost pure $\sim$GeV $\nu_{\mu}$ beam from accelerators
with more than hundreds kilometer long baseline.
From the MINOS\cite{MINOSe} experiment, it was reported that $\sin^{2}2\theta_{13} < 0.12$.  

T2K (Tokai to Kamioka)\cite{T2KLOI} is a new long-baseline neutrino-oscillation experiment in Japan.
A high-intensity neutrino beam from the J-PARC Main Ring is shot toward
the Super-Kamiokande (SK) detector, 295~km away.
It is claimed \cite{T2KLOI} that the T2K experiment is sensitive
for $\sin^{2}2\theta_{13} > 0.006$ in 90\% C.L.

\section{T2K neutrino beam line and detectors}

\begin{center}
\begin{figure}[h]
\hspace{2pc}
\includegraphics[width=35pc]{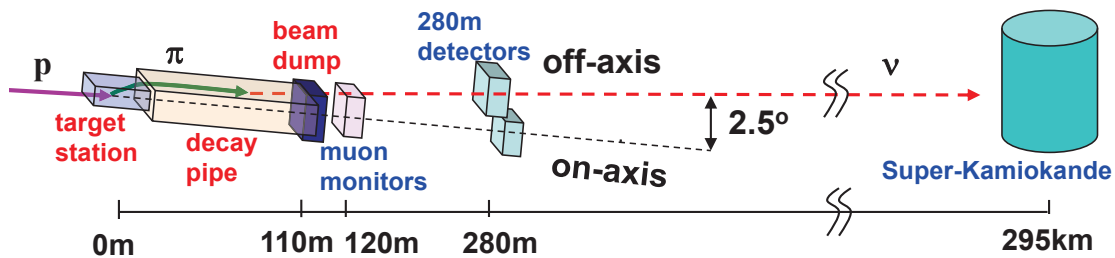}
\caption{\label{fig:concept}A schematic view of the T2K neutrino beamline and detectors. Beamline components are indicated
by the words in red, and the detectors are indicated by words in blue.}
\end{figure}
\end{center}

A schematic view of the T2K neutrino beamline and detector components are shown in Fig.\ref{fig:concept}.
Details on the beamline and the detector components have been comprehensively discussed in \cite{t2knim}.
In this article, some of the important features are presented below.

\subsection{Neutrino beam line}
The proton beam for the T2K experiment is extracted from the J-PARC 30-GeV Main Ring proton synchrotron.
This beam is delivered to the target station where it strikes the carbon target, resulting in the production of
pions and kaons; the pions and kaons are bent toward the direction of SK by a magnetic field
produced by magnetic horns.
Neutrinos are produced as decay products of pions and kaons.
All particles except neutrinos and high-energy ($>$ 5~GeV) muons are absorbed by a beam dump
located 110~m downstream of the target.

The most important feature of the T2K neutrino beamline is the off-axis beam,
which was originally proposed in \cite{offaxis}.
The details of the concept of the off-axis angle in T2K
have been reported in \cite{crimea}.
When the beamline construction was started,
the off-axis angle was tunable\cite{crimea} because
the optimum  off-axis angle was unknown.
The value of $\Delta m^{2}_{23}$ from other experiments\cite{SKatm,MINOS2006} was considered,
and the off-axis angle was accordingly adjusted to be 2.5$^{\circ}$ in 2007.
The corresponding peak energy of the neutrino flux is approximately 600~MeV,
and the oscillation study is the most sensitive around $\Delta m^{2}_{23} = 2.5\times 10^{-3}{\rm eV^{2}}$.

\subsection{Muon monitors}
Two types of muon monitors are installed downstream of the beam dump:
the Ionization Chamber and the Semiconductor Array.
They can measure the intensity distribution of muons that escape from the beam dump.
Since these muons are mainly produced along with neutrinos from the $\pi^{+}$ decays,
the beam center of muons is in accordance with the beam center of neutrinos. 
The position of the beam center can be monitored on a bunch-by-bunch basis,
within a 3-cm resolution, from the peak of the muon intensity distribution.
This position resolution corresponds to a 0.25-mrad accuracy of beam direction.

\subsection{Near detectors}
The near detectors were constructed in the underground experimental hall at a depth of
33.5~m, with a diameter of 17.5~m, and 280~m downstream from the target.
Two detectors were installed: the on-axis detector in the direction of the
neutrino beam center and the off-axis detector in the direction of SK.
\medskip

\begin{figure}[t!]
\center{{\includegraphics[height=6.0cm]{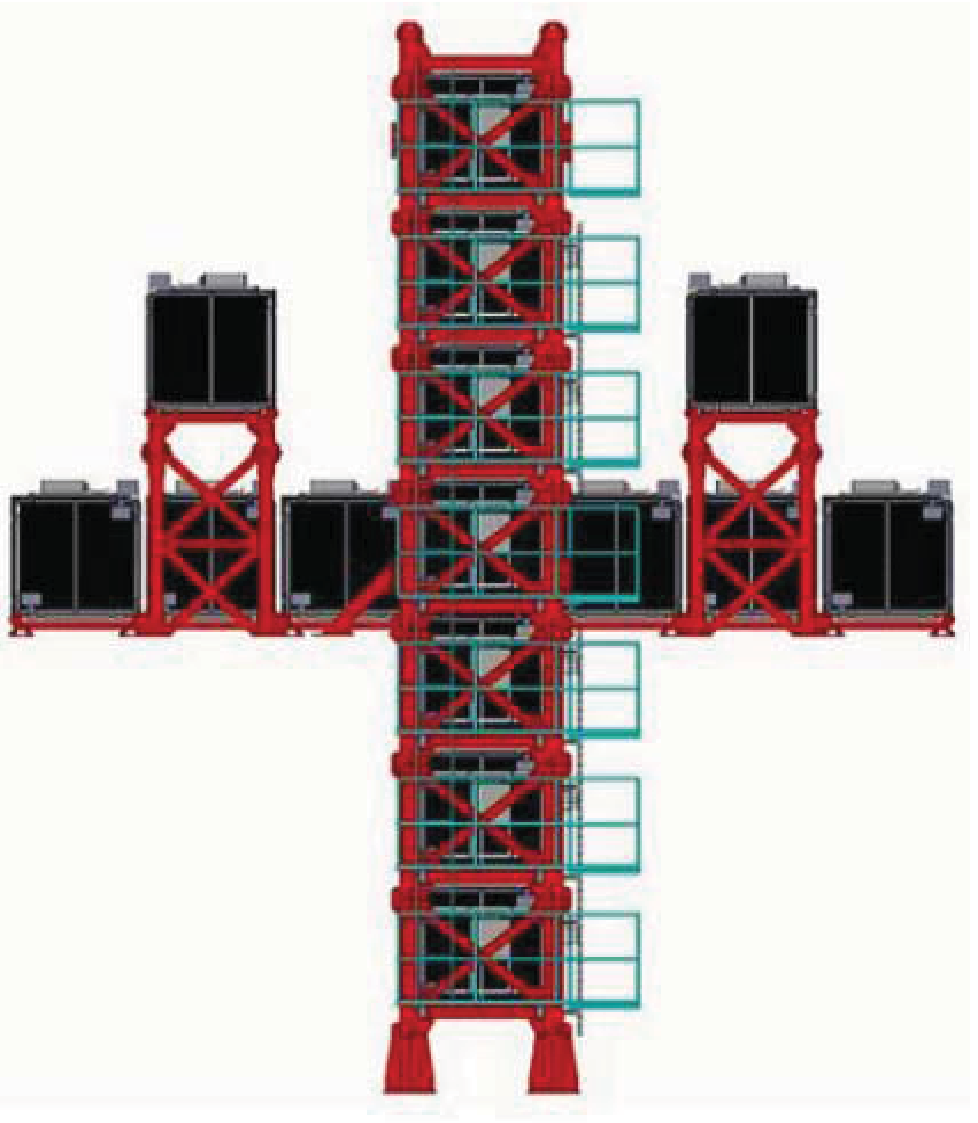}}
\hskip 2.0cm
        {{\includegraphics[height=6.0cm]{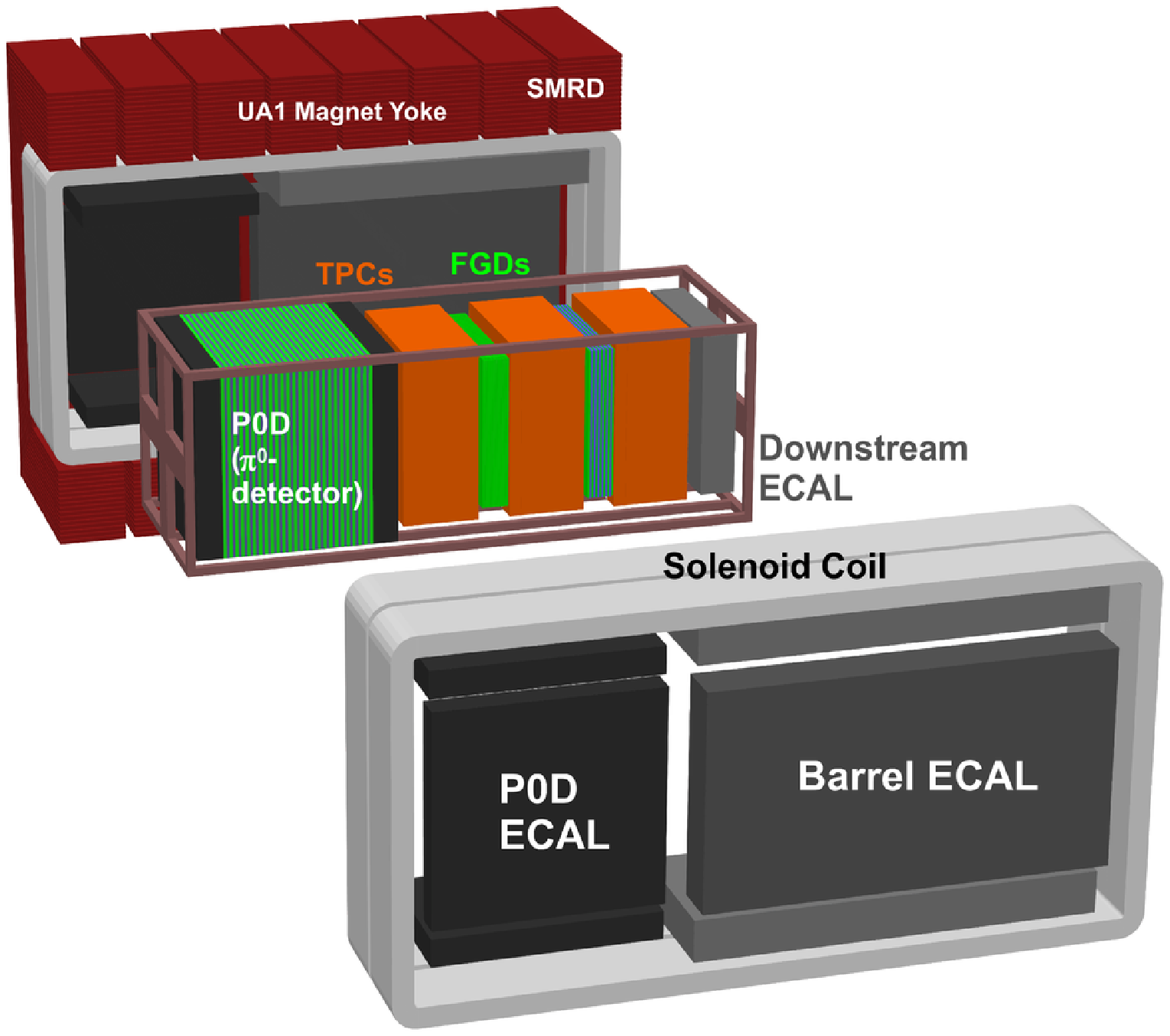}\vskip 0.5cm}}
}
\center{
\caption{\label{fig:near}
Schematic view of the INGRID on-axis detector (left) and
the ND280 off-axis detector (right).
}}
\end{figure}

The on-axis detector, referred to as the INGRID detector, consists of
16 1m$\times$1m$\times$1m cubic modules, as shown in Fig.\ref{fig:near}~(left).
Each module is a \char'134 sandwich" of
11 scintillator layers and 10 iron layers.
They are surrounded by 4 veto planes.
The modules are arranged in such a way that seven are horizontally aligned, seven are
vertically aligned, and two are placed in an off-diagonal fashion.
The neutrino beam center can be measured from the horizontal/vertical distribution
of the neutrino event rate.

\medskip

The off-axis detector is named ND280. The purposes of the ND280 detector are
\begin{itemize}
\item measurement of the energy spectrum of muon neutrinos in the SK direction,
\item measurement of fraction of electron neutrinos, and
\item study of neutrino interactions, especially, neutral current $\pi^{0}$, which is background
for the $\nu_{e}$ appearance signal.
\end{itemize}
\noindent
A schematic view of the ND280 detector is shown in Fig.\ref{fig:near}~(right).
All the detector components except SMRD are placed inside a 0.2~T magnetic field produced by the recycled UA1
magnet from CERN.
\noindent
The Pi zero Detector (P0D) is a subdetector component placed in the upstream inside
the magnet. It is a \char'134 sandwich" of a water target, scintillator planes, and lead plates.
It is customized for measurement of neutral $\pi^{0}$ production. $\gamma$-rays from
$\pi^{0}\rightarrow 2\gamma$ are converted to electromagnetic showers using lead
plates and are detected by scintillators.

Three Time Projection Chambers (TPCs) and 2 Fine Grained Detectors (FGDs) are placed downstream
of the P0D.
The TPCs can measure the momentum of muons from the curvature of the muon track in the magnetic field.
The momentum resolution is better than 10\% at 1~GeV.
FGDs consist of scintillator bars. They provide the target material for
neutrino interactions
and are optimized for detecting the proton recoils.
By combining the TPCs and FGDs, the energy spectrum of $\nu_{\mu}$
can be precisely measured from charged current quasi-elastic (CCQE) neutrino interactions.

Other detector components, Side Muon Range Detector (SMRD) and Electromagnetic Calorimeter (ECAL)
are installed surrounding the P0D and TPC/FGD.
They are reported in detail in \cite{t2knim}.

\subsection{Super-Kamiokande}
The far detector, SK is a 50-kton water Cherenkov detector\cite{SKdetector}.
It is located underground at a depth of 1000~m at the Kamioka mine, Japan.
The distance of SK from J-PARC is 295~km.
In the inner detector (ID), 22.5~kton fiducial volume
is viewed using 11129 20-inch diameter PMTs.
The outer detector (OD), surrounding the ID, is also a water Cherenkov detector.
It is used to veto particles that enter or exit the ID.
SK started its operations in April 1996. After an accident involving PMTs in 2001,
it was fully recovered by June 2006.

The most important characteristic of SK as the far detector 
of the T2K experiment is its particle identification capability between muons and electrons.
This particle identification directly relates to identification
between parent $\nu_{\mu}$ and $\nu_{e}$.
The principles of particle identification have been reported in \cite{Kasuga}.
It was verified that the $\mu$/$e$ misidentification probability is less than 1\%\cite{Kasuga}. 

\begin{center}
\begin{figure}[b]
\includegraphics[width=36pc]{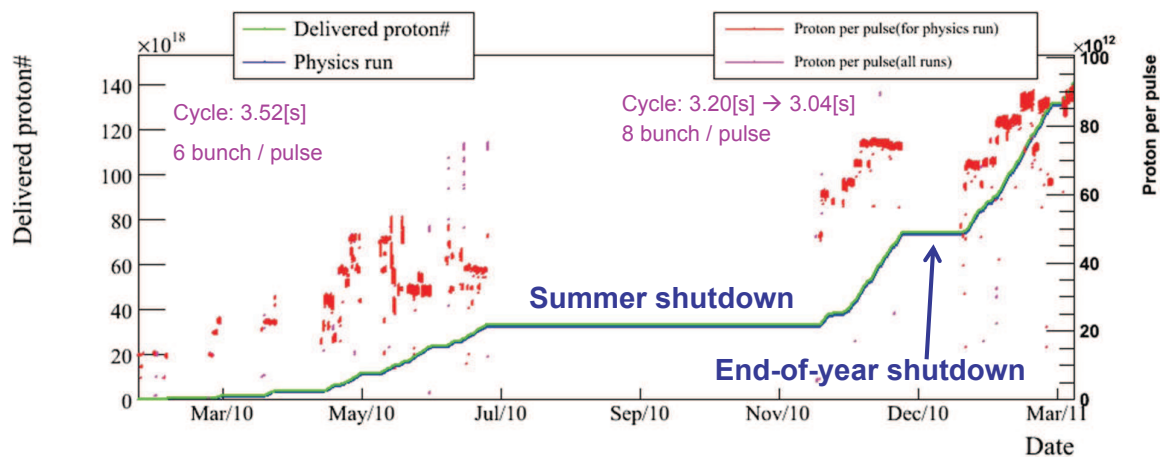}
\caption{\label{fig:beamhistory}History of primary proton beam intensity in the T2K experiment.
Red dots indicate the number of protons per pulse, and the scale is given in the right vertical axis.
The green solid line indicate the integrated number of delivered protons from the beginning of the experiment.
The scale is given in the left vertical axis.
}
\end{figure}
\end{center}

\section{Status of data collection}

\begin{center}
\begin{figure}[t]
\center{
        {\includegraphics[height=4.0cm]{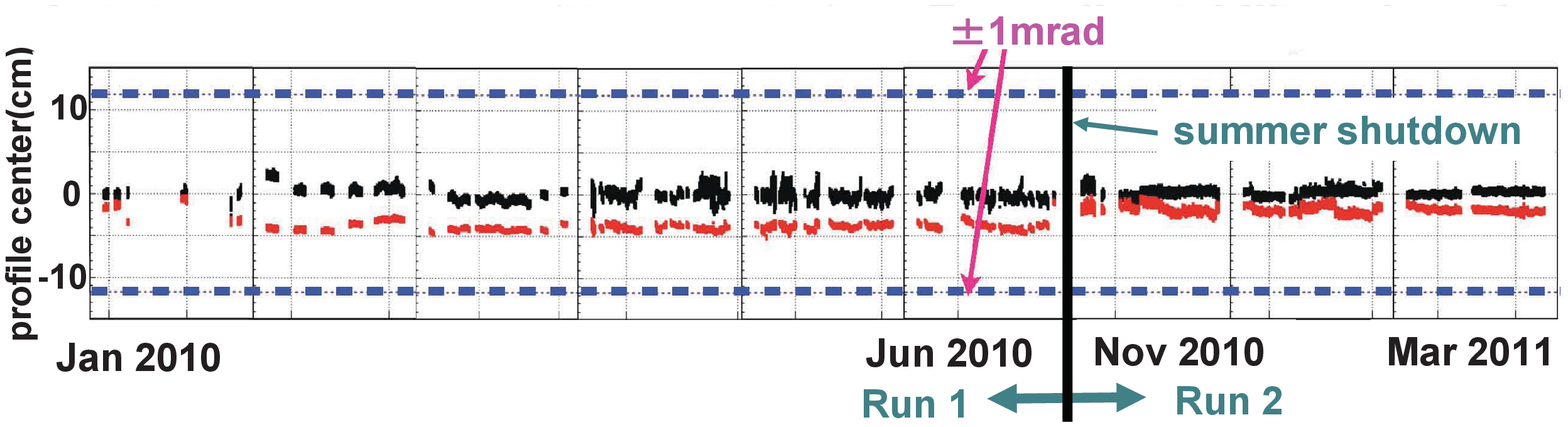}}
}
\caption{\label{fig:mumoncenter}Muon beam direction measured using muon
monitors downstream of the beam dump.
The black dots represent the horizontal profile center,
and the red dots represent the vertical profile center.
The beam direction is well controlled and is stable within 1~mrad.
}
\end{figure}
\end{center}

\begin{center}
\begin{figure}[t]
\center{
        {\includegraphics[height=7.0cm]{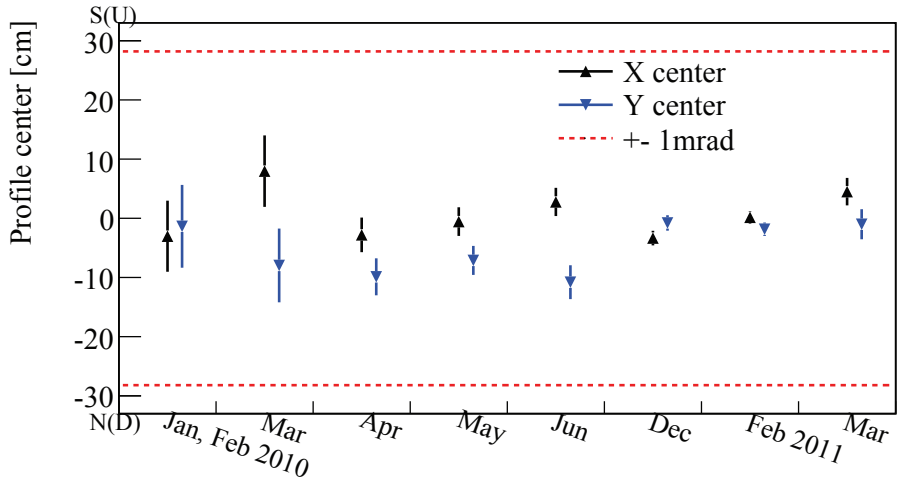}}
}
\caption{\label{fig:ingrid-data}Neutrino beam direction measured using the INGRID detector
at 280~m downstream from the target\cite{T2Kapp}.
The beam direction is well controlled and is stable within 1~mrad.
}
\end{figure}
\end{center}

The first proton beam was delivered to the target station on April 23, 2009.
Generation of neutrinos was confirmed by the signal of the muon monitors.
The first neutrino event in INGRID was detected on November 22, 2009.
The physics data collection started in January 2010, and the first neutrino event in SK
was observed on February 24, 2010.

The history of proton beam delivery is shown in Fig.\ref{fig:beamhistory}.
The beam duration of the T2K beam is $\sim$5~$\mu$s.
In early 2010, the beam cycle was 3.52~s, and each pulse contained 6 bunches.
This was changed to a beam cycle of 3.04~s, and 8 bunches per pulse in order to
increase the beam power. 
The number of protons per pulse (ppp) was improved from
$\sim 2\times10^{13}$~ppp to  $\sim 9\times10^{13}$~ppp.
The last beam power corresponds to about 150~kW.
This value is about 1/5 of the design value, 750~kW.
We have accumulated 1.43$\times$10$^{20}$ protons on target (pot) until March 11, 2011.
This is about 2\% of our goal, i.e., $\sim 78\times$10$^{20}$ pot, which can be attained by
the beam operation for 5~years with 750~kW.

The stability of the beam direction is continuously monitored using the muon monitors
(Fig.\ref{fig:mumoncenter}) and the INGRID detector (Fig.\ref{fig:ingrid-data}).
The beam direction is stable and well controlled within an accuracy of $\pm$1~mrad.
Note that if the beam direction is changed by 1~mrad, the neutrino flux at SK is changed
by 2\% at $E_{\nu}=0.5\sim 0.7$GeV. Therefore, the accuracy of the beam direction
is satisfactory.

\section{Electron neutrino appearance analysis}

The possible $\nu_{e}$ appearance signal was searched for using the 1.43$\times$10$^{20}$ pot data\cite{T2Kapp}.
The results of the event selection and expected background are summarized in Table~\ref{tab:reduction}.
Details of the event selection are given below.
\subsection{Event selection}

Beam-related fully contained (FC) events are selected to satisfy the following 3 conditions.
\begin{description}
\item[~~~~1-1)] The total energy deposit in the ID to be larger than 30~MeV
\item[~~~~1-2)] No OD activity
\item[~~~~1-3)] The SK event time to be within the timing range from -2 to 10~$\mu$s around the beam trigger time
\end{description}
The third condition involves an examination of the correlations
between the GPS time recorded in Tokai and that recorded in SK.
After these three conditions, 121 FC events are selected.
The accidental contamination from non-beam related events is estimated to be 0.023.

\begin{table}[b!]
\caption{
Number of selected events after each condition.
T2K data from 1.43$\times$10$^{20}$~pot, and expectations
from a Monte Carlo simulation with systematic errors.
The assumptions in the simulation
are $\Delta m^{2}_{23}=2.4\times 10^{-3}{\rm eV^{2}}$, $\sin^{2}2\theta_{23}=1$,
and $\sin^{2}2\theta_{23}=0$.
}
\smallskip
\begin{center}
\begin{tabular}{lcc}
\hline
\hline
                          &     Data    &    ~~~~~~~~~~  Expected     \\
\hline
1)Fully Contained (FC)                     &    121     &  ~~~~~~~~~~ 109    \\
2)Fully Contained Fiducial Volume (FCFV)      &     88     &  ~~~~~~~~~~ 74.1    \\
3)Single ring $e$-like ($E_{{\rm vis}} > 100{\rm MeV}$)  ~~~~~~~~~~~~~   &  ~7      & ~~~~~~~~~~ $5.8\pm 2.2$ \\
4)Additional $\nu_{e}$ cut               &               ~6       &   ~~~~~~~~~~   $1.5\pm 0.3$ \\
\hline
\hline
\end{tabular}
\end{center}
\label{tab:reduction}
\end{table}

A spatial reconstruction algorithm is applied to the 121 FC events,
and the following condition is applied. 
\begin{description}
\item[~~~~2)] The vertex position in the 22.5~kton of the fiducial volume
\end{description}
After this condition, 88 events are selected as fully contained fiducial volume (FCFV)
events.

A ring counting algorithm and the particle identification algorithm are 
applied to the 88 FCFV events.
The events are categorized as
single-ring $e$-like events, single-ring $\mu$-like events, and multi-ring
events. The following events are selected.
\begin{description}
\item[~~~~3)] Single ring $e$-like events with visible energy larger than 100~MeV
\end{description}
As presented in Table~\ref{tab:reduction}, 7 single-ring $e$-like events
remain. Note that the above analysis procedures are almost the same
as the well-established atmospheric neutrino analysis procedures\cite{SKatm}, except
1-3).

\begin{figure}[b!]
\center{{\includegraphics[height=6.5cm]{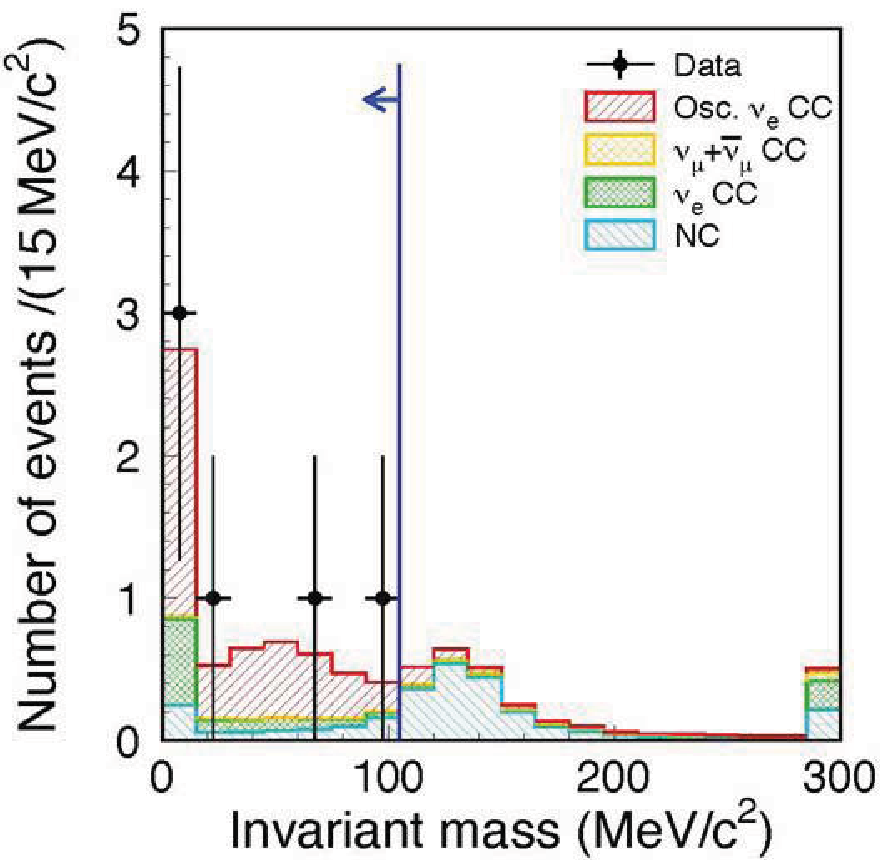}}
\hskip 1.0cm
        {{\includegraphics[height=6.5cm]{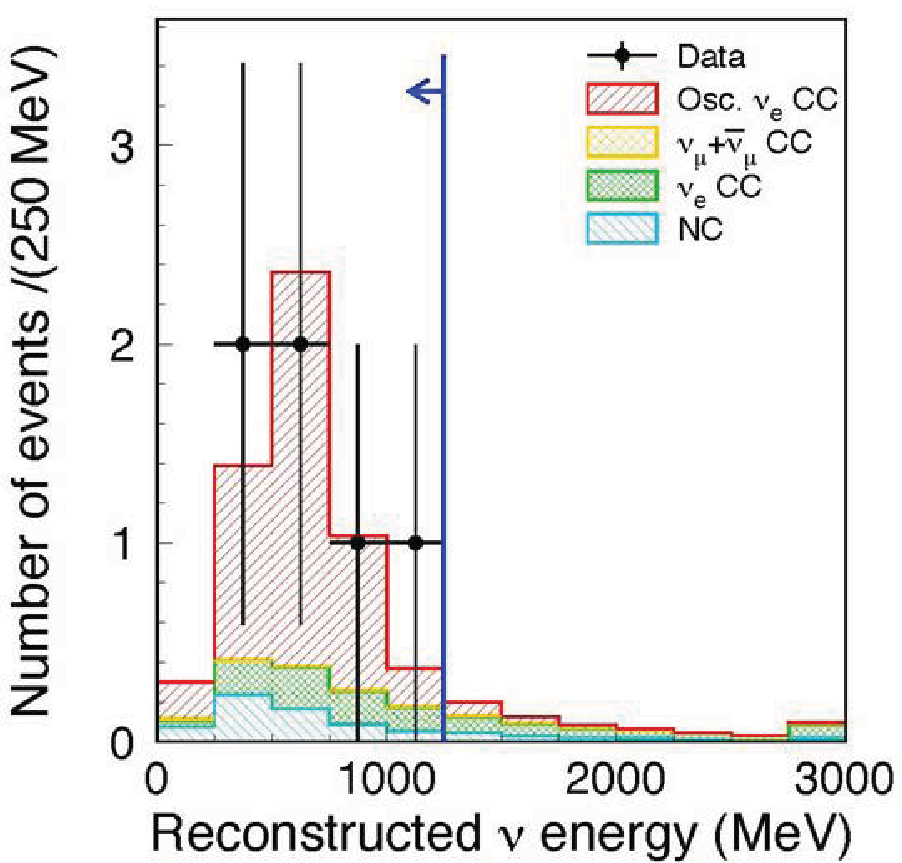}\vskip 1cm}}
}
\center{
\caption{\label{fig:additional}
(left)~Distribution of invariant mass $M_{inv}$ when each event is forced to be reconstructed into two
rings\cite{T2Kapp}. The data are shown using points with error bars (statistical only). The MC predictions
are in shaded histograms, corresponding to the oscillated $\nu_{e}$ CC signal and various background sources
for $\sin^{2}2\theta_{13}=0.1$. 
The vertical line shows the applied cut at 105 MeV/c$^{2}$.
(right)~Reconstructed neutrino energy spectrum of the events that satisfy all conditions of
$\nu_{e}$ appearance signal selection\cite{T2Kapp}. The vertical line
shows the applied cut at 1250 MeV.
}}
\end{figure}

To enhance the purity of the $\nu_{e}$ appearance signal,
3 additional conditions are required. They are
\begin{description}
\item[~~~~4-1)] No delayed-electron signal
\item[~~~~4-2)] Non-$\pi^{0}$-like event
\item[~~~~4-3)] Reconstructed neutrino energy to be less than 1250~MeV
\end{description}
\noindent
From condition 4-1), events with a delayed-electron signal are rejected because such a signal
indicates the generation of muons. One event is rejected considering this condition. 
Condition 4-2) is for suppression of misidentified $\pi^{0}$.
The reconstruction of the second ring is forced,
and the 2-ring invariant mass, $M_{inv}$, is calculated.
The $M_{inv}$ distribution is shown in Fig.\ref{fig:additional}~(left).
The condition $M_{inv}~<~105$~MeV is imposed.
No events are rejected considering condition 4-2).
From condition 4-3), the neutrino energy $E_{\nu}^{rec}$ is calculated by assuming quasi-elastic
kinematics. To suppress the contribution from the intrinsic $\nu_{e}$ component arising
primarily from kaon decays, the event is rejected if $E_{\nu}^{rec}~>~1250$~MeV.
As shown in Fig.\ref{fig:additional}~(right), no events are rejected by condition 4-3).
Finally, 6 events remain after considering all selection criteria.

\subsection{Expected background and its systematic error}
The expected background for $\nu_{e}$ appearance is carefully calculated through
simulations of neutrino productions and their interactions.

From the interactions between the primary proton beam and the carbon target,
secondary pions and kaons are generated.
Pions and kaons propagate in the secondary beamline.
Neutrinos are produced as decay products of pions and kaons.
All of these neutrino production processes are simulated using
standard simulation programs, FLUKA\cite{FLUKA1,FLUKA2},
GEANT3\cite{GEANT3}, and GCALOR\cite{GCALOR}.
To reduce the systematic uncertainties in the pion and kaon production simulation,
some of the T2K members join CERN NA61 experiment: \char'134 Study of hadron
productions in hadron-nucleus and nucleus-nucleus collisions in CERN SPS".
Pion production data\cite{NA61} for a 30-GeV proton beam incident on a carbon target
was obtained with 5$\sim$10\% error and is employed in the simulation.
The estimated uncertainties of the intrinsic $\nu_{e}$ flux below 1~GeV is around 14\%

The neutrino interactions are simulated using the NEUT\cite{NEUT} neutrino interaction generator.
The GENIE\cite{GENIE} generator is also used for the evaluation of uncertainty due to the simulation program.
The present uncertainty of neutrino interaction comes from limited experimental data.
For example, the uncertainty of CCQE interactions relative to the CCQE total cross section
is energy dependent and is about 7\% at 500~MeV.

\begin{center}
\begin{figure}[b!]
\hspace{5pc}
\includegraphics[width=25pc]{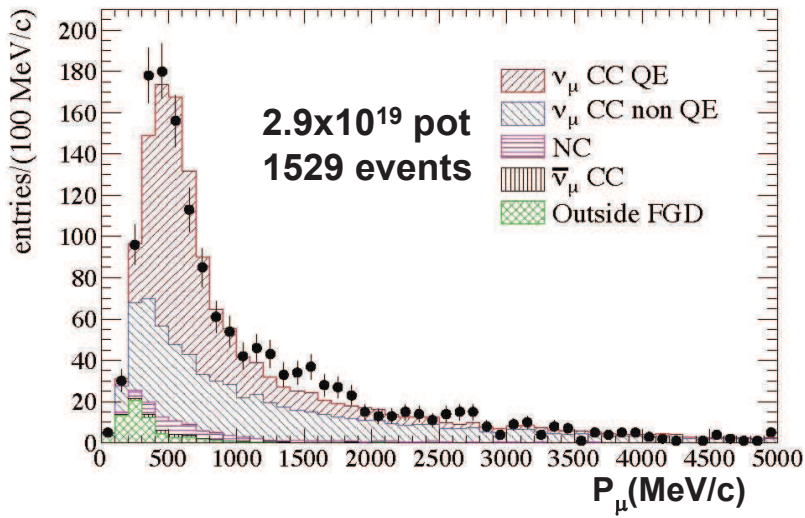}
\caption{\label{fig:nd280mom}Measured muon momentum of $\nu_{\mu}$ charged current candidates
reconstructed in the FGD target\cite{T2Kapp}. The
data are shown using points with error bars (statistical only),
and the MC predictions are represented by histograms.}
\end{figure}
\end{center}

\begin{center}
\begin{figure}[t]
\includegraphics[width=41pc]{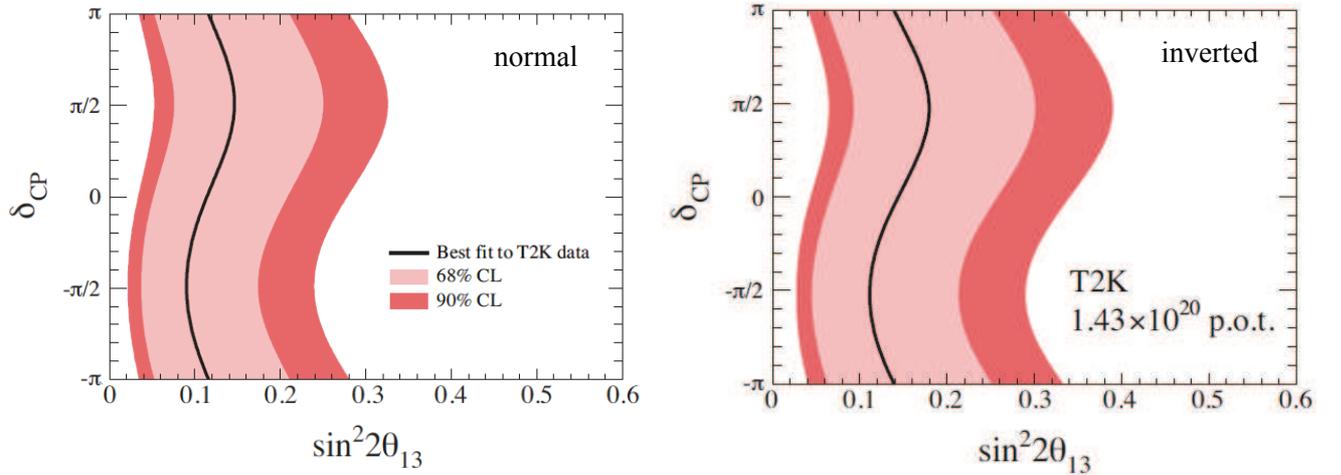}
\caption{\label{fig:contour}The 68\% and 90\% C.L. regions for $\sin^{2}2\theta_{13}$
for each value of $\delta_{{\rm CP}}$, consistent with
the observed number of events in the three-flavor oscillation case for normal (left) and inverted
(right) mass hierarchy\cite{T2Kapp}. The best fit values are indicated by solid lines.}
\end{figure}
\end{center}

To reduce the uncertainty on the absolute event rate,
the neutrino event rate at ND280 is used. The charged-current inclusive
muon events generated in FGD are selected and compared with
the simulation data. The agreement of the total event number
and its momentum distribution is excellent, as shown in Fig.\ref{fig:nd280mom}.
The ratio, $R_{{\rm data/MC}}$, is calculated to be 
$$R_{{\rm data/MC}} = 1.036 \pm 0.028 {\rm (stat.)} ^{+0.044}_{-0.037}{\rm (det.syst.)}\pm 0.038 {\rm (phys.syst.)}$$ 
\noindent
This factor is used as a correction for the normalization of the SK event rate.
After employing $R_{{\rm data/MC}}$ for the near/far cancellation,
the systematic uncertainty on an expected background
for the $\nu_{e}$ appearance signal due to
the uncertainty of the neutrino interaction is 14\%.

The event selection algorithm presented in the previous subsection was also
applied to the events generated by the simulation.
The results are summarized in Table~\ref{tab:reduction}.
The final background events for the $\nu_{e}$ appearance signal are 1.5$\pm$0.3(syst.).
Of these, the contribution from the beam intrinsic $\nu_{e}$ is 0.8 and that
from neutral current interaction is 0.6.

\subsection{Results of electron neutrino appearance and discussion}
\begin{table}[b]
\caption{
Comparisons of the $\nu_{e}$ appearance analysis between T2K
and MINOS
}
\smallskip
\begin{center}
\begin{tabular}{lll}
\hline
\hline
                          &     T2K    &      MINOS      \\
\hline
Proton on target ($\times$10$^{20}$)        &    1.43      &   8.2     \\
Expected number of background events        & 1.5$\pm$0.3       & 49.5$\pm$2.8(syst.)$\pm$7.0(stat)    \\
Observed number of                          &  6     &  62 \\
~~~~~~ possible $\nu_{e}$ appearance events &        &     \\
Excess  &  2.5$\sigma$   &  1.7$\sigma$ \\
Constraints on $\sin^{2}2\theta_{13}$ &   $\sin^{2}2\theta_{13}=0.11$   &   $\sin^{2}2\theta_{13}=0.04$ \\
~~~~~~~~(Normal hierarchy, $\delta_{{\rm CP}}=0$)      &   ~~~~{\scriptsize (Best fit)}  &   ~~~~{\scriptsize (Best fit)} \\
                                      &   $0.03 < \sin^{2}2\theta_{13} < 0.28$  &  $\sin^{2}2\theta_{13} < 0.12$\\
                                      &   ~~~~{\scriptsize (90\% C.L. interval)}  &   ~~~~{\scriptsize (90\% C.L. interval)} \\
Constraints on $\sin^{2}2\theta_{13}$ &   $\sin^{2}2\theta_{13}=0.14$  &   $\sin^{2}2\theta_{13}=0.08$ \\
~~~~~~~~(Inverted hierarchy,  $\delta_{{\rm CP}}=0$)   &   ~~~~{\scriptsize (Best fit)}  &   ~~~~{\scriptsize (Best fit)} \\
                                      &   $0.04 < \sin^{2}2\theta_{13} < 0.34$  &  $\sin^{2}2\theta_{13} < 0.19$ \\
                                      &   ~~~~{\scriptsize (90\% C.L. interval)}  &   ~~~~{\scriptsize (90\% C.L. interval)} \\
\hline
\hline
\end{tabular}
\end{center}
\label{tab:T2KMINOS}
\end{table}

T2K observed 6 possible $\nu_{e}$ appearance signal for the expected background of 1.5$\pm$0.3,
as reported in the previous subsections. The chance probability that 6 or more events are observed
for the expectation of $1.5\pm 0.3$ is calculated
to be  $7\times 10^{-3}$, equivalent to a 2.5$\sigma$ significance.
Although this statistical significance is still far from discovery or even from evidence,
it should be reported that it is an {\bf indication} of $\nu_{e}$ appearance.

The constraints on $\sin^{2}2\theta_{13}$ for assumed $\delta_{{\rm CP}}$ are shown in
Fig.\ref{fig:contour}. If normal hierarchy $(m^{2}_{3} >> m^{2}_{2} > m^{2}_{1})$ and
$\delta_{{\rm CP}}=0$
are assumed,
the best fit is $\sin^{2}2\theta_{13}=0.11$, and the 90\% C.L. interval is
$0.03 < \sin^{2}2\theta_{13} < 0.28$.
On the other hand, 
in the case of inverted hierarchy $(m^{2}_{2} > m^{2}_{1} >> m^{2}_{3})$
and $\delta_{{\rm CP}}=0$,
the best fit value is $\sin^{2}2\theta_{13}=0.14$ and the 90\% C.L. interval is
$0.04 < \sin^{2}2\theta_{13} < 0.34$.
These results do not contradict with the 90\% C.L. upper limits
by the CHOOZ\cite{CHOOZ} experiment and the MINOS\cite{MINOSe} experiment.

Just after the ICPP-II Istanbul conference, the MINOS collaboration
announced new preliminary results\cite{MINOSnew} on
the $\nu_{e}$ appearance analysis with 8.2$\times$10$^{20}$ pot data.
They observed 62 events whereas the expected background is 
$49.5\pm 2.8({\rm syst})\pm 7.0({\rm stat})$. The excess is 1.7$\sigma$.
Results from T2K and MINOS are summarized in Table~\ref{tab:T2KMINOS}.
The signal-to-background ratio in T2K is quite larger than that in MINOS.
This is the effect of the off-axis beam in T2K.
The peak of the neutrino energy is adjusted to the oscillation maximum,
and high-energy neutrinos that produce a neutral current $\pi^{0}$ background
are suppressed.

At present, 3 reactor experiments\cite{DC, Reno, Daya} are under construction or just started.
The indication of non-zero $\sin^{2}2\theta_{13}$ will be examined by the
reactor experiments and/or T2K additional data within a couple of years.

\section{The earthquake}

On March 11, 2011, Japan encountered a 9.0 Magnitude earthquake.
It was the 4th largest earthquake in the world since 1900.
The earthquake and followed tsunamis resulted in the death of more than 15000 people.
In addition, the Fukushima nuclear power stations, which was damaged by the tsunamis,
were found to have crucial radiation leaks.

J-PARC is located about 260~km from the seismic center, and about 100~km
from the Fukushima nuclear power stations.
No one at J-PARC was injured by the earthquake and tsunamis, but some of the
accelerator and beamline components were damaged.
After the radiation leak from Fukushima,
the radiation dose in the J-PARC area was considerably higher than the normal value.
However, the radiation dose has already returned to the normal
value, and recoveries of the accelerator/beamline components are in progress.
We will resume the J-PARC operation in December 2011 and restart the T2K data-taking
as soon as possible.

\end{document}